\documentclass[%
 amsmath,amssymb,
 reprint,%
]{revtex4-1}

\usepackage[T1]{fontenc}
\usepackage{mathptmx}

\usepackage[english]{babel}
\usepackage{graphicx}

\usepackage{dcolumn}
\usepackage{epstopdf}
\usepackage{bm}
\usepackage{mhchem}

\newcommand{\AO}{Pt/PZT$_{AO}$}
\newcommand{\BO}{Pt/PZT$_{BO_2}$}
\newcommand{\AOI}{Pt/PZT$_{AO}$/Pt}
\newcommand{\BOI}{Pt/PZT$_{BO_2}$/Pt}

\begin{document}
\preprint{AIP/123-QED}

\title{Interfacial phenomena in nanocapacitors with multifunctional oxides}

\author{A. V.\ Kimmel}
\affiliation{CIC nanoGUNE, Tolosa Hiribidea, 76, San Sebastian, 20018, Spain} 
\email{a.kimmel@nanogune.eu}

\begin{abstract}

The analysis of the structure, chemical stability, electronic and ferroelectric properties  of the interfaces between Pt(001) and 
PbZrTiO$_3$(001) have been performed with $ab$~ $initio$~ methods. 
We show that the chemical environment plays a critical role in determining 
the interfacial reconstruction and charge redistribution at the metal/oxide interfaces. 
We demonstrate that the difference in interfacial bonds formed at the Pt/PZT interfaces 
with (TiZr)O$_2$ - and PbO- termination of PZT essentially defines the effectiveness of the screening, 
and ease of polarisation switching in PZT-based capacitors.
%
The imperfect screening in \BOI~ capacitors is caused by  
strong interfacial bonds formed at the \BO~ interface that is  accompanied by 
the suppressed polarisation of PZT film.
In contrast, the capacitors with PbO-terminated  PZT show a negligible depolarising field, and high polarisation, 
which is the consequence of weak bonds formed at the Pt/PZT interfaces. 
The latter also causes a higher switching barrier than that in the former system.
\end{abstract}

\maketitle

%
%
\begin{section}{INTRODUCTION}
\indent Stable ferroelectric phases in nanometer-thick films are of great interest for  ultra-high density and  ferroelectric field
effect transistors (FeFETs), thus, the properties of ferroelectrics in ultrathin films has been under intense theoretical
and experimental investigation ~\cite{ Scott2011,  Zhang2018, Zubko2014, Tagantsev2006, Al-Saidi2010, JJones2016, Nozaka98, Chen2011, Bucur2017, Au1998}.

Lead zirconate titanate Pb(Zr$_{1-x}$Ti$_x$)O$_3$ (PZT) is one of the most commonly used ferroelectric materials due to its small coercive field, large polarisation, relatively high Curie temperature, and excellent piezoelectric response.~
PZT is a disordered solid solution ABO$_3$ perovskite, with Pb atoms occupying the A-site, and Ti and Zr cations randomly arranged among the B-sites.  
The material exhibits a rich phase diagram with the transition region, $x$=0.42-0.52,  known as the morphotropic phase boundary (MPB), where it exhibits its highest piezoelectric response  ~\cite{Noheda1999, Glazer2014}.

PZT has found a wide range of applications in piezoelectric sensors, actuators, and non-volatile memories, where
the interface between the metal electrode and a functional oxide plays an important role in the performance of the device ~\cite{ Al-Saidi2010, Nozaka98, Chen2011, Bucur2017, JJones2016, Au1998, Setter2006, Lou2009, Tagantsev2001}, 
For thin PZT  films, in particular, the cycling stability, the imprint performance and the leakage current strongly depend on the interfaces between  electrode  material and ferroelectric film ~\cite{JJones2016, Tagantsev2001, Lou2009, Han2010}.

%
%
However, the reliable and precise data on the interfacial properties, whether at the structural, or chemical level, is often difficult to obtain  from experiment. 
In the interfacial region the local chemical and electrostatic environment  significantly differs from that of  parent materials, 
and the description of the interface in terms of bulk parameters is unjustified. 


In this work, we use density functional theory
calculations to provide an insight into the  properties of  PZT(001)/Pt (001) interfaces. 
We study interfacial reconstruction, the nature of interfacial bonds, and interfacial charge redistribution.
We  characterise the stability of the ferroelectric phase in  Pt/PZT/Pt metal/oxide nano-capacitors with respect to the size and  termination of PZT. 
We show that the chemical bonding at the interface plays a critical role in determining the effectiveness
of the screening, ease of polarisation switching, and stability of ferroelectric state in perovskite-based capacitors.

\end{section}

\begin{section}{METHODS}
 
We carried out our first principles calculations with the CASTEP plane-wave code ~\cite{CASTEP} 
and ultrasoft pseudopotentials ~\cite{Vanderbilt_PRB_1990}. We chose the Wu-Cohen 
generalised gradient approximation density functional ~\cite{Wu_Cohen_GGA} because it is known
to reproduce the structural properties of perovskites with high accuracy. 

A 500 eV plane-wave cutoff together with 0.025 \AA$^{-1}$ density of $k$-points  Monkhorst-Pack mesh  have been used for calculations. The convergence of the total energy and forces per atoms were 0.005 eV and  0.05 eV/\AA, respectively.

\begin{figure} [htbp] 
\begin{center} 
\includegraphics[width=2in]{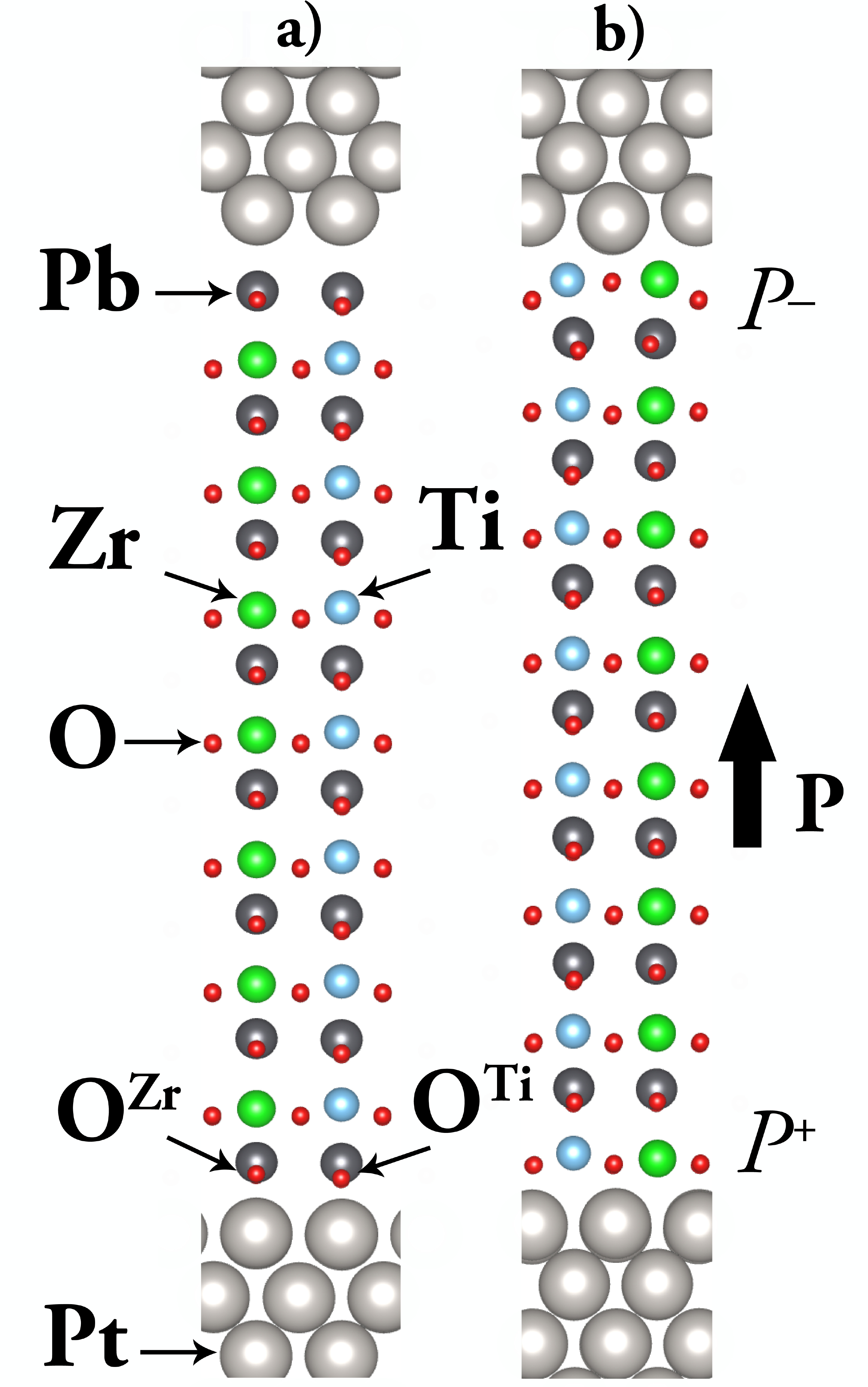}
\caption{ 
Schematic representation of modelled Pt/PZT capacitor  with
a) PbO- and b) TiZrO$_2$  - terminations of PZT. 
The direction of polar axis is perpendicular to the interface. 
Due to symmetry breaking there are two inequivalent interfaces further denoted as $P^+$ and $P^-$.
} 
\label{fig-setup} 
\end{center} 
\end{figure} 

We model MPB composition of PZT with $x$=0.5 using one of the stable $P4mm$ phases found by  DFT methods ~\cite{Bogdanov2016}.  
This  phase is characterised by the lattice parameters of $a$=5.625 \AA~  and  $c$=4.261 \AA,  and exhibits an alternating columnar arrangement of B-site cations  along the polar axis (Fig.~\ref{fig-setup}a, b). 

We have modelled a PZT(001) slab terminated with PbO, and (TiZr)O$_2$ atomic layers, further denoted as AO-,  and BO$_2$ as followed to standard definition of perovskite formula ABO$_3$, where A-site cations are Pb, while B-site cations are Ti and Zr. 
Further, we use \AO~ and \BO~ notations for the Pt/PZT interfaces with AO and BO$_2$ terminations of PZT, respectively.
Due to  symmetry breaking, each simulated capacitor system is characterised by two electrostatically inequivalent interfaces further denoted as $P^+$ and $P^-$ (see Fig.~\ref{fig-setup}).

To reveal the effects of the thickness of ferroelectrics in Pt/PZT/Pt capacitor geometry 
we have varied the thickness of PZT, $n$, from 1.5 (0.06 nm) to 7.5 (3.55 nm) unit cells.

We model metal electrodes using Pt slabs with a 3.998 \AA ~fcc lattice parameter.  
To achieve the  convergence of electrostatic potential inside the metal we used ten  layers  of the metal electrode overall. 

To simulate the effect of the mechanical boundary conditions due to the strain 
imposed by the substrate, the in-plane lattice constant of PZT was fixed to the theoretical equilibrium 
lattice constant of bulk Pt, thus, the tensile strain was  of 0.6~\%.
PZT atoms were allowed to relax in all the directions, while only three metal layers in vicinity of the interface were allowed to relax.

The  Pt/PZT/Pt nanocapacitors  were simulated by using a supercell approximation with imposed periodic boundary conditions. 
A (1x1) periodicity of the supercell perpendicular to the interface is assumed. 
This set up gives the electrical short-circuit condition as well as an electrode/perovskite super-lattice geometry.  
Thus, the formula describing the periodically repeated supercell is Pt$_{5}$/PZT$_n$/Pt$_5$, $n$=1.5 ,..., 7.5.

We calculate the rumpling parameter, $\nu$, which provides indirect understanding of a dipole bearing by an atomic layer,
as an averaged deviation of cations ($M$) and anions ($O$) in atomic layer, $i$, along the polar axis, $z$ ~\cite{Stengel2011}:
$$
\nu_z=\sum_i \delta^z(M_i) - \delta^z(O_i).          (1)
$$

The band offset at a metal/insulator interface was calculated using a spatially resolved projected density of states (PDOS)
by defining the location of the band edges deep in the insulating region, with the Fermi level of the metal taken as a reference ~\cite{Stengel2011}: 

$$
\rho(i,E) = \sum_{nk} \int_BZ  <i|\psi_{nk}>|^2 \delta (E-E_{nk}) dk,           (3)
$$
where $|i>$ is a normalised function localised in space around the region of interest, and  $\rho(r, E)$ is the local density of states.

The planar-averaged electron charge density normal to the interface was calculated as:
$$
\rho(z)=1/S_{xy} \int \int ~\rho(x, y, z) dx dy,             (4)
$$
where $S_{xy}$ is a cross-section area of the system, $\rho(x, y, z)$ is three-dimensional charge density and $\rho(z)$  is its planar averaged derivative.

The planar averaged potential was calculated across the capacitor as:
$$
V(z)=1/S_{xy}\int \int V(x, y, z) dx dy,           (5)
$$
where $V(x, y, z)$ is three-dimensional electrostatic potential, $S_{xy}$ is the cross-section of ferroelectric.
The planar averaged potential, $V(z)$, exhibits fast oscillations correlated with the atomic arrangement in the material. In order to 
take the information related to interfaces we have used a macroscopically averaged potential ~\cite{Stengel2011}:
$$
<V(z)>=\int V(z) H(z-z^\prime) dz^\prime,    (6)
$$
where $H(z)$ is the Heaviside step-function taken with the period of the PZT lattice parameter.

\end{section}


\begin{section}{RESULTS}

\begin{subsection}{Chemical stability and atomic relaxation}

To characterise the interfacial binding energy density $\sigma$ we calculated the difference between 
the total energy of the modelling system, $E_{tot}$, and the sum of the energy of the PZT slab ($E_{PZT}$) 
and the Pt slab ($E_{Pt}$) normalised to the surface area, $S$, of the interface (since our system has two interfaces we include a factor of 2 into the equation):
$$
\sigma= [( E_{PZT}^{slab} + E_{Pt}) - E_{PZT-Pt} ]/ 2S.
$$

We have found that both, \AO~ and \BO, systems  are chemically stable.  
Calculated  $\sigma$ values exhibit a variation with the thickness of PZT. 
However, the increase of $n$ leads to a rapid convergence of  $\sigma$ 
to 0.175 eV/\AA$^2$~ and 0.252 eV/\AA$^2$ for \AO~ and  \BO~ systems, respectivly.  
Thus, it is energetically advantageous to form a Pt/PZT interface, although, the \BO~ system is 94.5 meV/\AA$^2$  more stable than \AO.
The values of the binding energy density are in good agreement with previous $ab$~$initio$~ calculations of similar Pt/PZT interfaces calculated with open-circuit boundary conditions ~\cite{JJones2016}.

We have found a  strong difference in the interfacial structure of Pt/PZT related to different chemical environment. 
The \AO~ interface is characterised by a close contact of the PbO atomic plane with the surface of the metal electrode.
%
%
One may distinguish different types of oxygen species at the \AO~ interface (See Fig. ~\ref{fig-setup}):  
the oxygen in  Pt-O-Zr bonds, further denoted as O$^{Zr}$, and the one in Pt-O-Ti bonds  denoted as O$^{Ti}$.
We show further that despite the chemical similarity of Ti and Zr cations in PZT \cite{Pontes2012}, 
the interfacial bonds  and the charge transfer at the interface with Pt are different for these species.

With  the increase of $n$, the  length of interfacial  Pt-O$^{Ti}$ and Pt-O$^{Zr}$ bonds 
converges to  2.1 \AA~and  1.99 \AA~  for $P^+$,  and to 2.35 \AA~and 2.7 \AA~ for $P^-$, respectively.
The electrostatic asymmetry also affects the second neighbours to the interface: 
for the $P^-$ interface the short Ti-O and Zr-O bonds are  0.1 \AA~ shorter in comparison to  their bulk values 
(1.91 \AA~ and 1.75 \AA~ for the short Zr-O and Ti-O bonds, respectively), 
while the long Ti-O and Zr-O bonds at the $P^+$ interface  were elongated by 0.4 \AA~ (2.22 \AA~ and 2.5 \AA~ for long Zr-O and Ti-O bonds, respectively). 

Such a  difference in bond lengths occurs due 
%
the relative proximity of Pb and O ions to the surface of the electrode. The latter
affects the electrostatic repulsion between cations dominated by the attraction between  O and Pt ions.
Such an uncompensated repulsion, as we will later see, leads to the formation of weak Pt-O bonds at the $P^+$ interface.

The \BO~ interface also  shows $P^-$ and $P^+$ asymmetry with respect to the interfacial bonds.
Indeed, the  $P^+$ interface exhibits relatively short interfacial Pt-O bonds of  2.12 \AA.
However, the $P^-$ interface relaxes with two types of  Pt-O  interfacial bonds: long  (2.4 \AA) and short (2.12 \AA). 
At the $P^-$ the B-site cations  locate closer to Pt cations than O ions, thus, the repulsion between 
positive species is partly balanced by the Pt-O attraction.

We have also observed  rumpling of the first metal layer  of 0.1 \AA, which rapidly decays away from the interface.

The analysis of the Crystal Orbital Hamilton Populations curves related to interfacial bonds demonstrates a strong bonding character of interfacial Pt-O bonds at \BO~ interface. Meanwhile, these bonds exhibit a very weak, even non-bonding character at the $P^-$ side of the \AO~interface  (See Supplementary Information).

%
%

We have found that interfacial bonds formation induces the rotation of oxygen octahedra at the interfacial layers.
The rotation of  octahedra is  characterised with the tilt angle,  $\Theta$ (See Fig. \ref{fig_OR}).
With  increasing of  PZT thickness  a rapid decrease from 8$^{\circ}$ to 0.5$^{\circ}$ is observed. 
We found that the  tilt angles  of the octahedra based on  Zr and Ti cations  exhibit different signs, demonstrating  an antiferro-distortive nature of the  reconstruction at the Pt/PZT interfaces.
The tilt angles are bigger at the \AO~ interface, while 
the P$^+$ interface of \BO~ shows negligible angles ($\Theta<$1$^o$) (See Fig. \ref{fig_OR}). 

A similar phenomenon has been observed  at STO/PTO and STO:Nb/PTO interfaces ~\cite{Zhang2018}.  Here, the  rotation of oxygen octahedra was observed in nearly zero-strain systems at open-circuit conditions. The authors have observed a long range octahedral rotation with the critical thickness of 4 nm, above which the rotation disappears.

We assume that the origin of the interfacial rotation of oxygen octahedra at the Pt/ZT/Pt interfaces is related to the minimisation of strain  and the depolarising field. While, at the \BO~ interface the octahedral rotation has a lesser amplitude due to the strong interfacial Pt-O bonds formed.

%

\begin{figure} [htbp] 
\begin{center} 
\includegraphics[width=3.in]{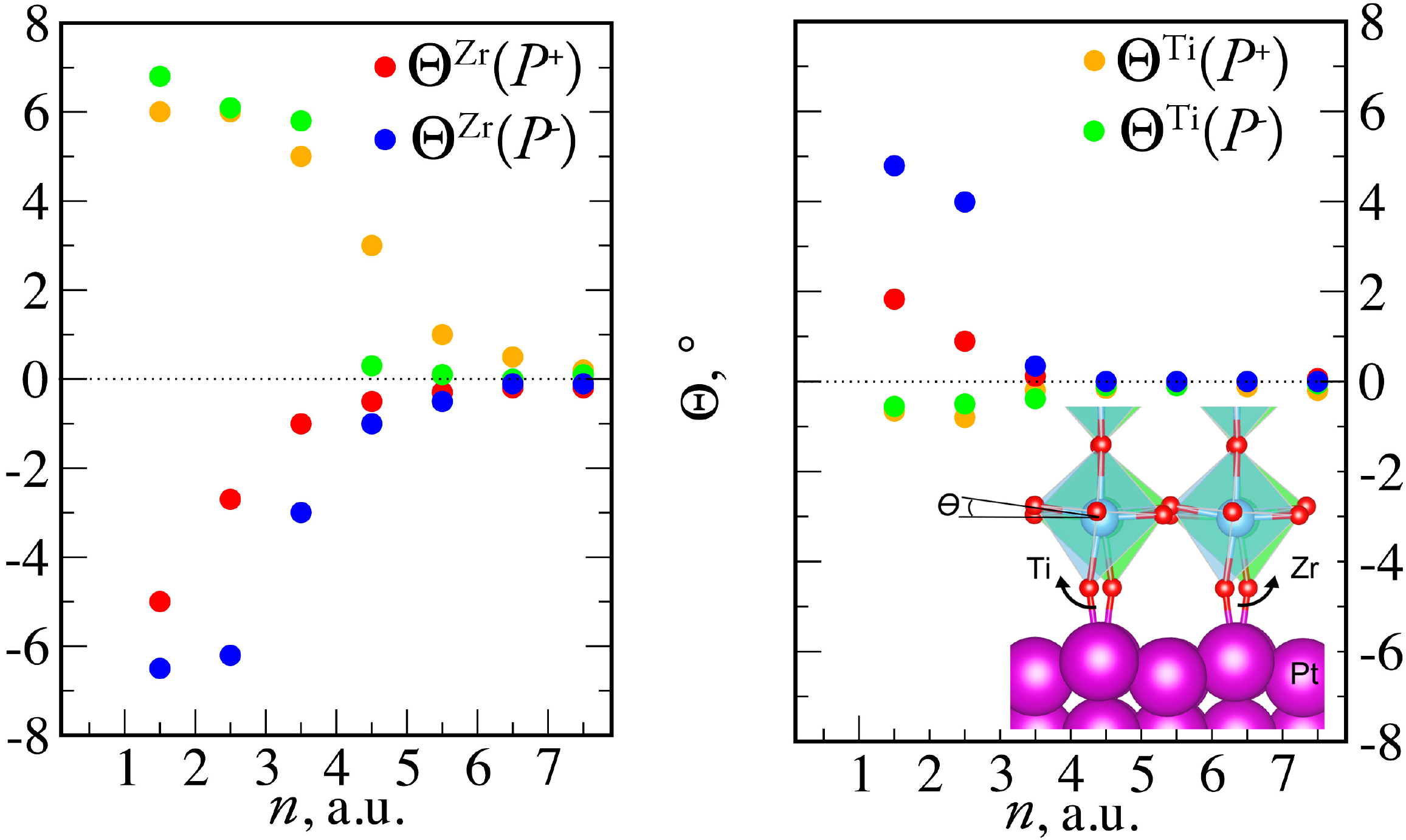}
\caption{ The dependence of the octahedral tilt, $\Theta$, 
 on the thickness, $n$,  at a) \AO~, b) \BO~ interfaces.} 
\label{fig_OR} 
\end{center} 
\end{figure} 

\end{subsection}

\begin{subsection}{Rumpling parameter}

In the capacitor geometry  the formation of the interfacial bonds may affect the polarisation with the possibility of suppression, or enhancement of polarisation in the ferroelectric material~\cite{Stengel2011}.
To characterise this, the rumpling parameter, $\nu$, for \AOI~ and \BOI~ capacitors has been calculated for different thicknesses of PZT (Fig. ~\ref{fig_rum}).

Below the critical thickness of $n_c$=4.5 u.c the rumpling parameter in \AOI~ exhibits  smaller values than that of the bulk  PZT (0.49 \AA~ and 0.36 \AA~for the PbO and TiZrO$_2$~atomic layers, respectively). 
With $n>n_c$ the rumpling reaches its bulk limit  for the PbO atomic layer, while $\nu$ for TiZrO$_2$ layers is  0.2 \AA~ lower than the bulk value.
The smallest system ($n$=1.5 u.c.) exhibits a negative rumpling at the $P^-$ side. This corresponds to the appearance of a head-to-head domain in PZT (See Fig. ~\ref{fig_rum}a).

Notably, the \AOI~ capacitor shows an enhancement of the rumpling parameter at the $P^+$ side up to 0.6 \AA,
because of  short Pt-O bonds formed, while the $P^-$ side exhibits reduced rumpling due to flattening of the interfacial  layer.

The overall rumpling parameter in the thin \BOI~ is significantly suppressed and reaches its bulk values  for the thicknesses above the critical one, $n_c$. 
The $P^+$ interface  including several adjacent layers  is characterised by a reduced rumpling parameter, 
while, the $P^-$ interface exhibits its slight increase, which converges to its bulk limit with $n>10$.

We assume that interfacial reconstruction affects \AO~ interfaces,  
making PbO atomic layers at $P^+$ rougher in comparison to bulk PZT.
Thus, we expect that polarisation in \AOI~ will be somehow similar to that in bulk PZT, while in \BOI~ capacitors we expect a suppressed polarisation due to flattening of the atomic layers.

This finding contradicts  the conclusion of ref.~\onlinecite{Stengel2009} that the perovskite materials with the  flatter surface displays a strong enhancement of the polar instability, while those that are significantly buckled shows its suppression. In the bulk $P4mm$ PZT the PbO atomic layer is more buckled than the BO$_2$ one. In the capacitor geometry the interfacial relaxation at the $P^+$ interface enhances the rumpling of the former, while notably reduces that of the latter.

\begin{figure} [htbp] 
\begin{center} 
\includegraphics[width=3.5in]{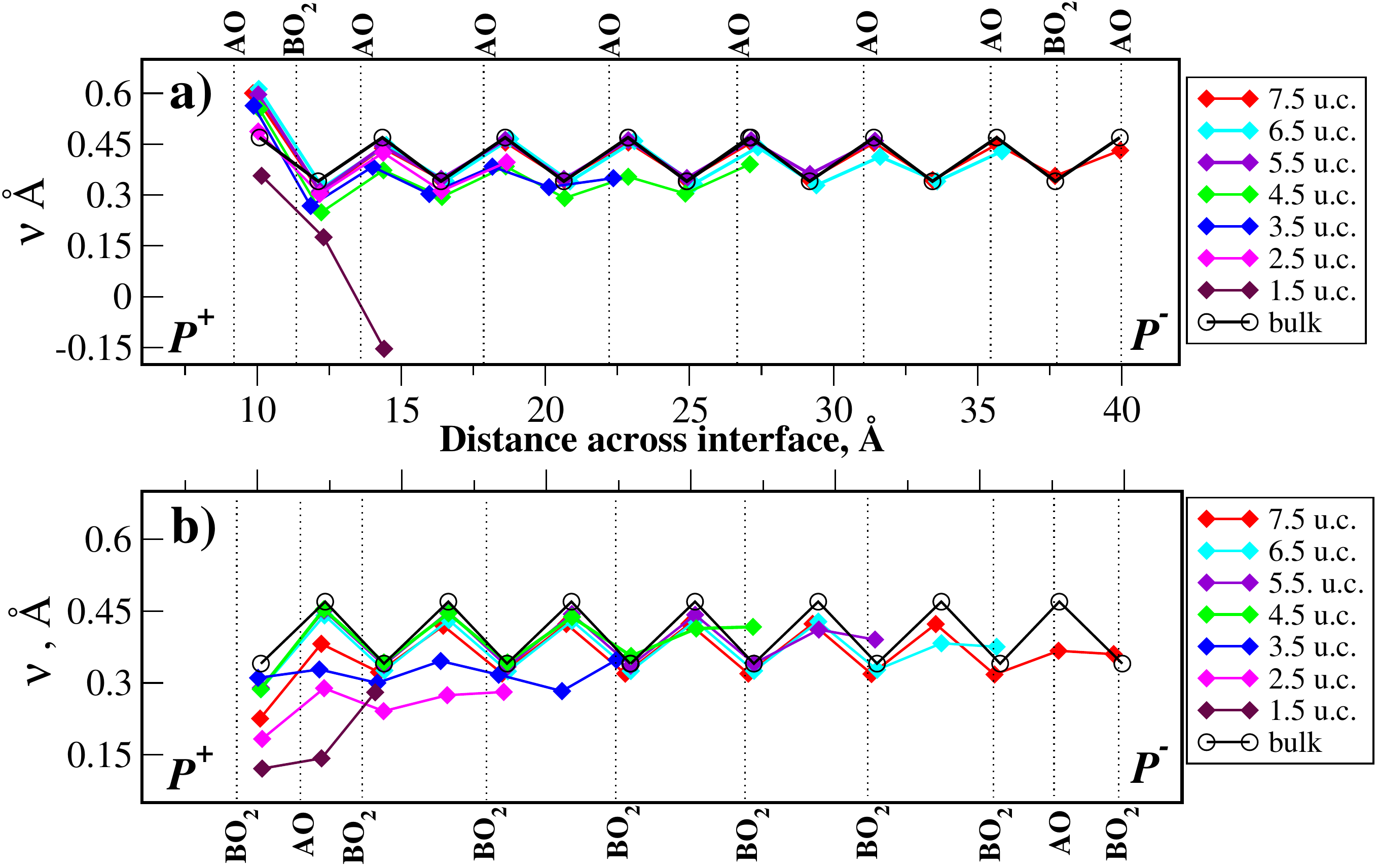}
\caption{ Dependence of the rumpling parameter, $\nu$ with respect to the slab thickness
and termination: a) \AO~ and b) \BO~systems. 
Open black circles corresponds to bulk PZT, while coloured ones corresponds to PZT in capacitor geometry with different thickness from 1.5 to 7.5 unit cells.
} 
\label{fig_rum} 
\end{center} 
\end{figure} 
\end{subsection}

%
%
\begin{subsection}{Switching}

The  interfacial reconstruction and bond formation may influence the switching characteristics of a capacitor 
because the process of switching requires a reversion of the polar state of ferroelectric, which
leads to a swap between $P^+$ and $P^-$ sides of the capacitor. 
Relative stability of the ferroelectric (FE) state versus the paraelectric (PE) state may be  affected by the interfacial reconstruction and important in understanding the switching properties of capacitors.

We have found that a PZT-based capacitor exhibits  a higher PE state energy than that of the FE state for all studied terminations and thicknesses (See further details in the Supplementary Information).
With $n>$4.5 u.c. the  switching barrier of \AOI~ converges to 0.41 eV per formula unit, while  that for \BOI~ gives a smaller value of 0.26 eV per formula unit. Both barriers are smaller than 0.451 eV barrier for bulk PZT.
The switching barrier for bulk PZT is expectedly high because it has been calculated for an ideal material free of defects (domain walls, metal contacts, vacancies). Clearly, the presence of interfaces affects the switching barrier.

We suggest that the switching barrier for the \BOI~  is smaller because  the interfacial layer is  affected by polarisation reversal in a small amount, i.e. there is no physical bond breaking at the interface, although bond lengths are modified.
In addition, the \BOI~ shows flatter atomic layers than \AOI, thus  less energy is required to revert the first system. 
In contrast, switching of \AOI~ requires the transformation of the interfaces with a strong Pt-O bonding (the $P^+$ side) to non-bonding ($P^-$), which consists in bond breaking.

\end{subsection}

%
%
\begin{subsection}{Charge density distribution}

\begin{figure} [htbp] 
\begin{center} 
\includegraphics[width=3.3in]{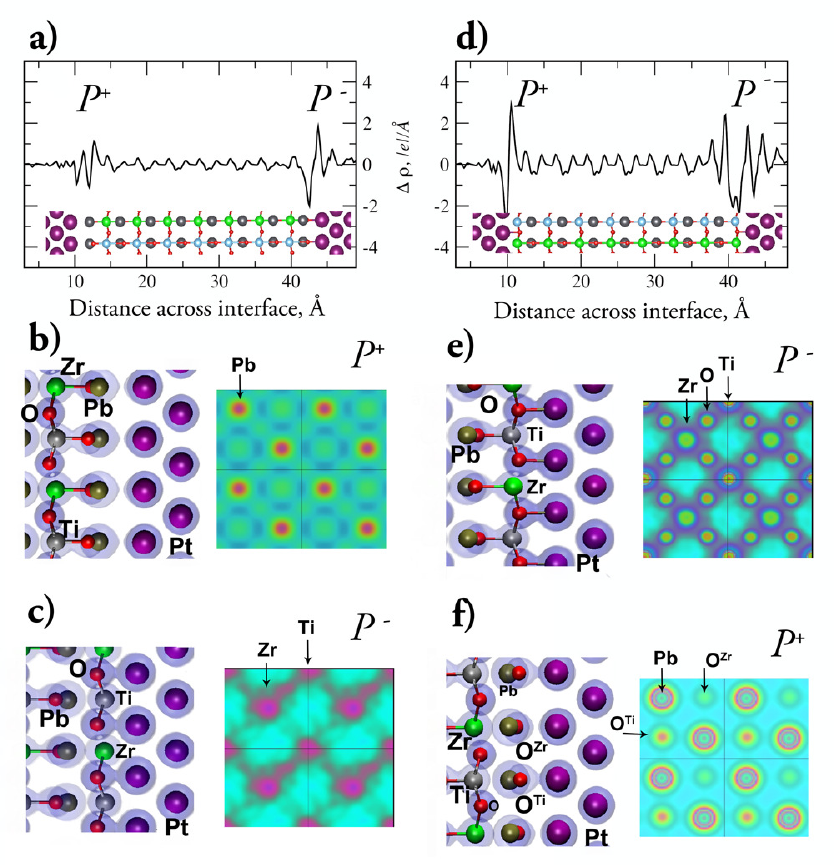}
\caption{
The planar average of charge density difference  of the  Pt/PZT$_{(7.5)}$ system and the 
sum of the isolated metal and perovskite densities for a) \AO~ and d) \BO.
The projections of electron charge density on
b) (xz) and (xy) projections of the charge density at $P^-$, 
c) (xz) and (xy) projections for P$^+$ interface of ~\AO. 
$P^{+}$ exhibits a larger population of the density originated from O$^{Ti}$, 
while $P^{-}$ shows homogeneous weak population.
e) (xz) and (xy) charge density projections for the P$^+$, 
f)  (xz) and (xy) projections for the P$^-$  interface of ~\BO. 
The $P^+$ interface shows  the charge density  localised at the interfacial bonds, 
while $P^-$ exhibits a lesser density at the interface with an asymmetry associated with O$^{Zr}$ species.
}
\label{fig_chg} 
\end{center} 
\end{figure}
The asymmetry  of interfacial bonds related to the different chemical (in a sense of AO- and BO$_2$-terminations), and the electrical  environment (i.e. the presence of $P^+$ and $P^-$ interfaces), indicates a strong charge redistribution at  Pt/PZT interfaces that would affect its electronic properties.
Further, we provide an insight into the redistribution of the charge density, $\rho$, in the Pt/PZT/Pt capacitors. 
For this, we  decompose the system into its perovskite and electrode components, and calculate charge densities separately  in respective frozen capacitor geometry. 
The difference between the charge densities of the Pt/PZT/Pt capacitor, and the sum of isolated Pt and PZT systems,  give the charge redistribution, $\Delta \rho$, caused by the interface (Fig. 4). The largest effect of the charge redistribution  is at the interface itself.
Inside the ferroelectric $\Delta \rho$ exhibits regular oscillations of a  small amplitude, indicating the long range effect of the density redistribution. Interesting, that the amplitude exhibited by the \BO~ is twice larger  than the that in the \AO. 

The analysis of the charge density at the crossover of  $P^-$ and $P^+$ interfaces
provides  an insight into the character of interfacial bonds (Fig. 4b, c, e, f).
%
In particular, the  \BO~  shows a dense homogeneous population of the charge density for both, $P^+$ and $P^-$, interfaces. This corresponds to strong chemical interactions followed by the charge transfer between Pt and the outermost layer of PZT. Despite the repulsion between cations at the $P^-$ interface, the oxygen states hybridise with Pt states. 

In contrast, the \AO~interface exhibits a weak  population of the charge density at the $P^-$ interface 
due to the repulsion between closely located positive charges of Pt and Pb cations.
We note the $P^+$ and $P^-$ asymmetry in the hybridisation between Pt and O states. 
The O$^{Ti}$ species exhibit charge density population at the Pt surface, 
while O$^{Zr}$ shows the density  localised in the ferroelectric.

To provide quantitative insight into the charge redistribution in a Pt/PZT$_{7.5}$/Pt system 
we have performed the analysis of the Bader charges~\cite{Bader09}.
For the reference, the Bader charges of oxygen species in bulk PZT vary as  -1.22 $|e|$ for species in Pb-O-Pb bonds, and as -1.05 $|e|$ and -1.28 $|e|$ for O$^{Ti}$ and O$^{Zr}$, respectively.

We found that interfacial Pt species show the charge modification of -0.3 and +0.3 $|e|$ in \AOI, 
and -0.2 and +0.2 $|e|$ in \BOI~   at the $P^+$ and $P^-$ interfaces, respectively.
 
At the $P^+$ side of Pt/PZT$_{AO}$ interface the  O$^{Ti}$ species become 0.07 $|e|$ more electronegative in comparison to bulk, 
while the O$^{Zr}$  loose 0.08 $|e|$.
However, at the $P^{-}$ the charge variation is more pronounced of -0.39 $|e|$ and +0.16 $|e|$ for the O$^{Ti}$ and O$^{Zr}$, respectively  (See notations in Fig. 4).

The anions at the \BO~ interfaces become less electronegative (see Fig. 4) with  charge reduction of 0.11 $|e|$ and 0.2 $|e|$ at the  $P^+$  and  $P^-$ interfaces, respectively. 
In addition,  Zr cations at  the \BO~interface became 0.02 $|e|$ and 0.06 $|e|$ more electropositive in comparison to the bulk at  the $P^+$ and $P^-$ sides, respectively. Meanwhile, Ti species   show the largest modification of the charge: at the $P^+$ interface Ti ion became 0.5 $|e|$ less electropositive, while at the $P^-$ it gains 0.154 $|e|$. 

The described interfacial charge redistribution  has a good correspondence with the projected charge density (Fig. 4b, c, e, f). 
The Bader charge analysis demonstrates that  Pt/PZT/Pt capacitors are characterised by a strong redistribution of the charge, which leads to  the formation of unequal local dipoles at the $P^+$ and $P^-$ interfaces, where  the \AO~ interface shows effectively larger formed dipoles  than that at the \BO~ interface. 

\end{subsection}

\begin{subsection}{Electronic properties}

For most practical applications, it is highly desirable for a capacitor to be insulating to $dc$ current.
An undesirable source of heating and power consumption is related to the transmission of electrons via non-zero conductivity or direct tunnelling.
%

\begin{figure} [htbp] 
\begin{center} 
\includegraphics[width=3in]{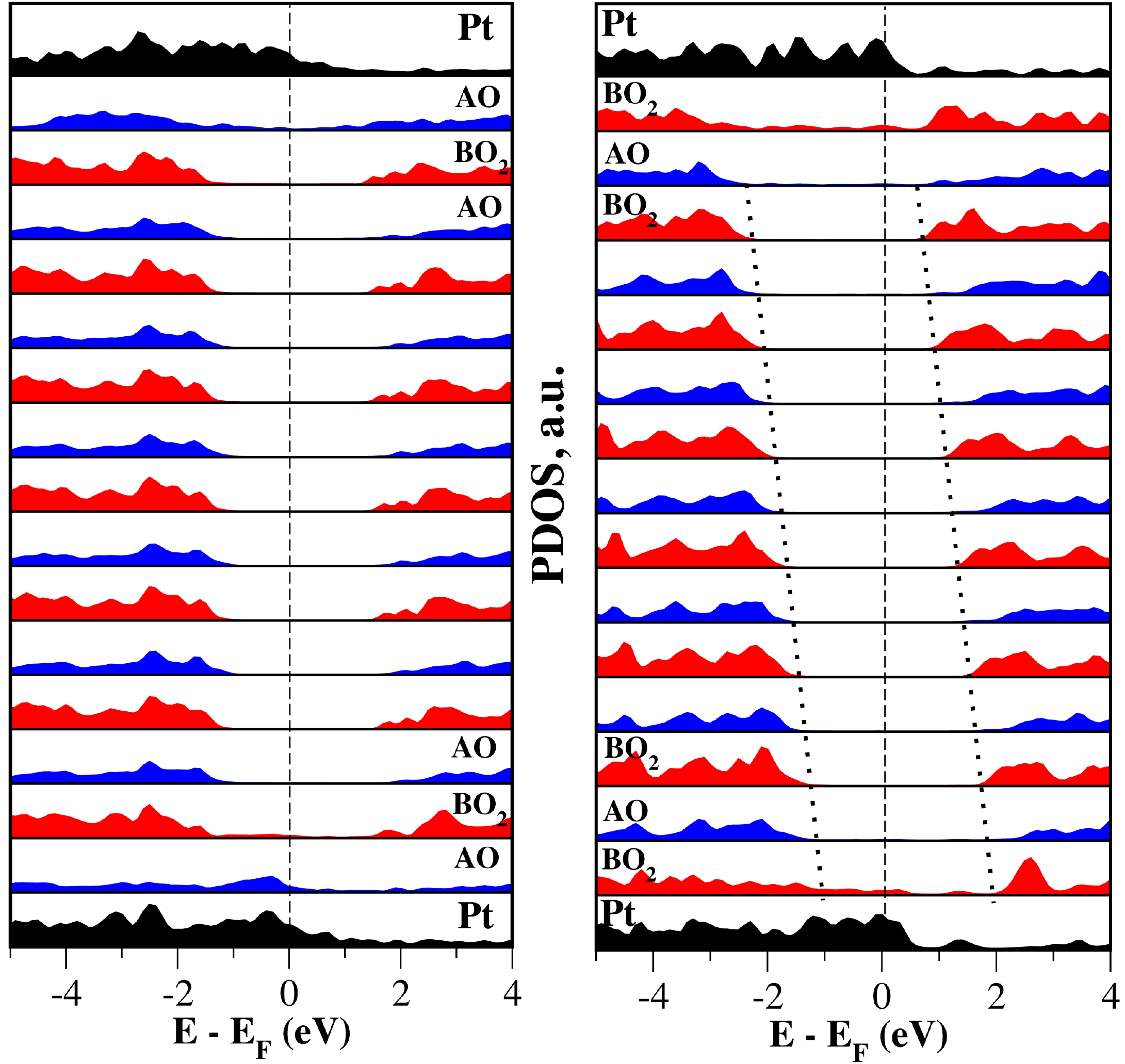}
\caption{
Projected Density of States  for atomic layers of (a) \AOI~ and (b) \BOI~ capacitors
with $n$=7.5 u.c. Red, blue and black areas correspond to AO-, BO$_2$-, and Pt atomic layers, respectively. 
The vertical broken line depicts the Femi level, while dotted lines depict top of the valence band and the bottom of the condition band of the PZT film. 
} 
\label{fig_dos8} 
\end{center} 
\end{figure} 

To provide an insight into the electronic properties of metal/oxide interfaces 
we have constructed PDOS  for \AOI~ and \BOI~ capacitors for the different PZT thicknesses.
We found that thin systems, with $n_c<$4.5 u.c., exhibit  populated electronic states  
at the Fermi level in the central layer of the system (See  Supplementary Information). 
The increase of  $n$ leads to an opening of the band gap, and its value  converges  
to 2.8 eV corresponding to the band gap of the  bulk PZT calculated with GGA level and expectedly 
underestimated in comparison to the  experimental value. 

In Fig. 5 the PDOS for Pt/PZT$_{7.5}$/Pt with AO- and BO$_2$-terminations of PZT are shown. 
One can see that the layers adjacent to the electrode interface exhibit states that cross the Fermi level, thus,  
the system is locally metallic.
For the \AOI~ capacitor in the middle of the PZT, the states at the Fermi level vanish, which implies that the system is locally insulating. 
The PDOS of the conduction and the valence bands converges quickly to the bulk curve when moving away from the interface. 

Furthermore, the PDOS of each layer of \BOI~ capacitor appears systematically shifted with respect to the neighbouring  layers, 
which is consistent with the presence of the depolarising field. The increase of the PZT thickness  
leads to  cross of the Fermi level with unoccupied levels of PZT. 
By extrapolating this straight line, we would see that it crosses the Fermi level at the thickness of $n$=9.5 u.c. 


\end{subsection}

\begin{subsection}{Screening properties}

Imperfect screening leads to the appearance of a depolarising field that  
reduces the dielectric response and can destabilise the single-domain ferroelectric state. 
Thus, an insight into  the interaction of ferroelectric and electrode materials is crucial for the device design.
We further analysed the planar averaged potential to tackle the variation of this property across the Pt/PZT/Pt  capacitor.
We plot the macroscopic-average electrostatic potential (Eq. 5), in Fig. 6 for \AOI~ and \BOI~ capacitors together with the potential in the freestanding PZT slabs  in frozen bulk geometry with a similar termination.

Due to the polarisation orientation, the $P^+$ and $P^-$ surfaces of the ferroelectric slab 
have different work functions causing  the potential drop across the  slab, $\Delta_1$, 
and the potential difference between two asymmetric vacuum potentials, $\Delta_2$ (See Fig. 6).
We found that the 
electronic screening properties of the PZT-based capacitors depend crucially on the local chemical environment.
Indeed,  the potential drop in  the AO-terminated  slab  ($\Delta_2-\Delta_1$= 0.17 V) is greater than that inside the \AOI~ capacitor $\Delta_3$=0.015 V. 
In the absence of electrodes, the depolarising field of 6.1 mV/\AA~ would bring the freestanding slab  into the paraelectric state,
while in the capacitor geometry  the  field is cancelled mostly due to metallic screening from the  electrodes that compensates the polarisation charge. Indeed, in \AOI~ the residual depolarising field inside the film is as small as 0.5 mV/\AA.

Similarly, for the BO$_2$-terminated slab we found a potential drop $\Delta_2$-$\Delta_1$  of 0.09 V, while the potential drop inside the BO$_2$-terminated capacitor is of 0.14 V much larger than that of the AO-terminated system (Fig. 6c, d).
The residual depolarising electric field in the \BOI~ capacitor is of 4.5 mV/ \AA, that produces a linearly increasing electrostatic potential on the film (Fig. 6b).
The existence of such a  depolarising field is caused by the imperfect screening at the interface ~\cite{Stengel2011, Rappe2005, Rappe2012}, which is a property of the interface as a whole (that is, the metal, the ferroelectric and the specific interface geometry). We assume the origin of the  depolarising field  is due to strong chemical bonds formed at the \BO~ interfaces. 
We therefore emphasise the importance of the chemical environment leading to the inequivalent  work functions  determining the ferroelectric behaviour of ultrathin films at this range of thicknesses.

\begin{figure} [htbp] 
\begin{center} 
\includegraphics[width=2.5in]{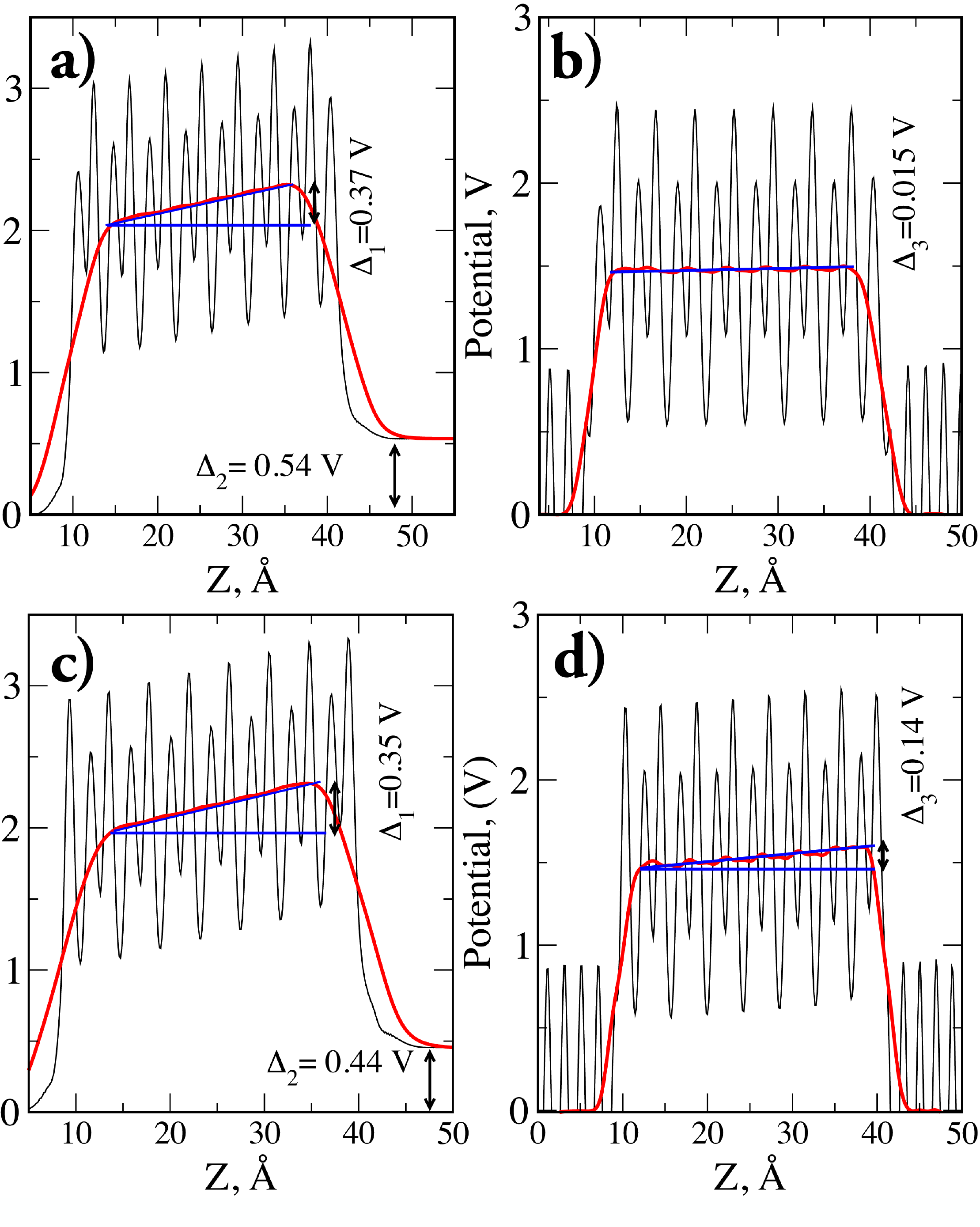}
\caption{Planar averaged electrostatic potential in 
a) the frozen bulk geometry freestanding AO-terminated PZT slab exhibits the drop of potential  $\Delta_1$=0.37 V, difference between the two asymptotic vacuum potentials is $\Delta_2$=0.54 V; 
b) freestanding BO$_2$-terminated PZT slab with $\Delta_1$=0.35 V,  vacuum potential drop is $\Delta_2$=0.44 V;
c) AO-terminated PZT slab in capacitor geometry exhibits drop of potential of $\Delta_3$= 0.14 V; 
d) BO$_2$-terminated PZT slab shows the drop of potential $\Delta_3$=0.090 V.
}

\label{fig-EP} 
\end{center} 
\end{figure} 

\end{subsection}

\end{section}


\begin{section}{SUMMARY}

In this work we  studied the structural and electronic properties of Pt/PZT interfaces with different, PbO- and TiZrO$_2$- terminations
in varied thickness capacitor geometry. 
We showed that the chemical environment at the interface plays a critical role in determining the interfacial bonds, promoting oxygen rotation and charge redistribution.
Our findings reveals that the \BO~ interface is ~94.5 meV/\AA~ more energetically stable
than \AO one.
%
This is related to strong interfacial bonds formed at the \BO~ interfaces,  
while the \AO~  is characterised by  weak interfacial bonds.
%
We expect a suppressed polarisation in the \BOI~capacitors due to 
a flattening  of its outermost atomic layers, while the \AOI~ capacitor shows an enhanced  rumpling parameter at the $P^-$ interface.

The ferroelectric state was found to be stable for all thicknesses of PZT films,  although,
thin capacitors with the thickness below 4.5 u.c. exhibit metal states in the gap.

The   Pt/PZT interfaces are characterised  by the strong charge redistribution leading to the formation of 
unequal local dipoles at the $P^+$ and $P^-$ interfaces, 
where  the \AO~ interface shows an effectively larger dipole formed than that at the \BO~ interface. 
%
The \AOI  capacitor exhibits a small depolarising field of 0.5 mV/\AA, while the \BOI capacitor 
shows a depolarising field ten times larger.
This is caused by the imperfect screening  due to strong interfacial bonds formed at the \BO~ interfaces 
and  suppressed rumpling of (ZrTiO)$_2$  outermost PZT layer.

We hope that our findings will provide a guidance to the synthesis  of nanoscale capacitors, and 
allow for a better control of epitaxial growth of PZT films.

\end{section}

{\bf Acknowledgements} 
{\small 
AK is supported by the European Union's Horizon 2020 research and innovation programme under the Marie Sklodowska-Curie 
grant agreement No 796781.
Author also acknowledges HPC-Europa programme for the access to ARCHER,  Barcelona Supercomputing Centre.
}

\bibliography{Interfacial_properties_Pt-PZT-FINAL.bib}

\begin{thebibliography}{26}%
\makeatletter
\providecommand \@ifxundefined [1]{%
 \@ifx{#1\undefined}
}%
\providecommand \@ifnum [1]{%
 \ifnum #1\expandafter \@firstoftwo
 \else \expandafter \@secondoftwo
 \fi
}%
\providecommand \@ifx [1]{%
 \ifx #1\expandafter \@firstoftwo
 \else \expandafter \@secondoftwo
 \fi
}%
\providecommand \natexlab [1]{#1}%
\providecommand \enquote  [1]{``#1''}%
\providecommand \bibnamefont  [1]{#1}%
\providecommand \bibfnamefont [1]{#1}%
\providecommand \citenamefont [1]{#1}%
\providecommand \href@noop [0]{\@secondoftwo}%
\providecommand \href [0]{\begingroup \@sanitize@url \@href}%
\providecommand \@href[1]{\@@startlink{#1}\@@href}%
\providecommand \@@href[1]{\endgroup#1\@@endlink}%
\providecommand \@sanitize@url [0]{\catcode `\\12\catcode `\$12\catcode
  `\&12\catcode `\#12\catcode `\^12\catcode `\_12\catcode `\%12\relax}%
\providecommand \@@startlink[1]{}%
\providecommand \@@endlink[0]{}%
\providecommand \url  [0]{\begingroup\@sanitize@url \@url }%
\providecommand \@url [1]{\endgroup\@href {#1}{\urlprefix }}%
\providecommand \urlprefix  [0]{URL }%
\providecommand \Eprint [0]{\href }%
\providecommand \doibase [0]{http://dx.doi.org/}%
\providecommand \selectlanguage [0]{\@gobble}%
\providecommand \bibinfo  [0]{\@secondoftwo}%
\providecommand \bibfield  [0]{\@secondoftwo}%
\providecommand \translation [1]{[#1]}%
\providecommand \BibitemOpen [0]{}%
\providecommand \bibitemStop [0]{}%
\providecommand \bibitemNoStop [0]{.\EOS\space}%
\providecommand \EOS [0]{\spacefactor3000\relax}%
\providecommand \BibitemShut  [1]{\csname bibitem#1\endcsname}%
\let\auto@bib@innerbib\@empty
\bibitem [{\citenamefont {Scott}(2011)}]{Scott2011}%
  \BibitemOpen
  \bibfield  {author} {\bibinfo {author} {\bibfnamefont {J.}~\bibnamefont
  {Scott}},\ }\href {\doibase 10.1146/annurev-matsci-062910-100341} {\bibfield
  {journal} {\bibinfo  {journal} {Annual Review of Materials Research}\
  }\textbf {\bibinfo {volume} {41}},\ \bibinfo {pages} {229} (\bibinfo {year}
  {2011})}\BibitemShut {NoStop}%
\bibitem [{\citenamefont {Zhang}\ \emph {et~al.}(2018)\citenamefont {Zhang},
  \citenamefont {Guo}, \citenamefont {Tang}, \citenamefont {Ma}, \citenamefont
  {Zhu}, \citenamefont {Wang}, \citenamefont {Li}, \citenamefont {Han},
  \citenamefont {Chen}, \citenamefont {Ma}, \citenamefont {Wu},\ and\
  \citenamefont {Ma}}]{Zhang2018}%
  \BibitemOpen
  \bibfield  {author} {\bibinfo {author} {\bibfnamefont {S.}~\bibnamefont
  {Zhang}}, \bibinfo {author} {\bibfnamefont {X.}~\bibnamefont {Guo}}, \bibinfo
  {author} {\bibfnamefont {Y.}~\bibnamefont {Tang}}, \bibinfo {author}
  {\bibfnamefont {D.}~\bibnamefont {Ma}}, \bibinfo {author} {\bibfnamefont
  {Y.}~\bibnamefont {Zhu}}, \bibinfo {author} {\bibfnamefont {Y.}~\bibnamefont
  {Wang}}, \bibinfo {author} {\bibfnamefont {S.}~\bibnamefont {Li}}, \bibinfo
  {author} {\bibfnamefont {M.}~\bibnamefont {Han}}, \bibinfo {author}
  {\bibfnamefont {D.}~\bibnamefont {Chen}}, \bibinfo {author} {\bibfnamefont
  {J.}~\bibnamefont {Ma}}, \bibinfo {author} {\bibfnamefont {B.}~\bibnamefont
  {Wu}}, \ and\ \bibinfo {author} {\bibfnamefont {X.}~\bibnamefont {Ma}},\
  }\href@noop {} {\bibfield  {journal} {\bibinfo  {journal} {ACS Nano}\
  }\textbf {\bibinfo {volume} {12(4)}},\ \bibinfo {pages} {3681} (\bibinfo
  {year} {2018})}\BibitemShut {NoStop}%
\bibitem [{\citenamefont {Lichtensteiger}\ \emph {et~al.}(2014)\citenamefont
  {Lichtensteiger}, \citenamefont {Fernandez-Pena}, \citenamefont {Weymann},
  \citenamefont {Zubko},\ and\ \citenamefont {Triscone}}]{Zubko2014}%
  \BibitemOpen
  \bibfield  {author} {\bibinfo {author} {\bibfnamefont {C.}~\bibnamefont
  {Lichtensteiger}}, \bibinfo {author} {\bibfnamefont {S.}~\bibnamefont
  {Fernandez-Pena}}, \bibinfo {author} {\bibfnamefont {C.}~\bibnamefont
  {Weymann}}, \bibinfo {author} {\bibfnamefont {P.}~\bibnamefont {Zubko}}, \
  and\ \bibinfo {author} {\bibfnamefont {J.-M.}\ \bibnamefont {Triscone}},\
  }\href@noop {} {\bibfield  {journal} {\bibinfo  {journal} {Nano Letters}\
  }\textbf {\bibinfo {volume} {14 (8)}},\ \bibinfo {pages} {4205} (\bibinfo
  {year} {2014})}\BibitemShut {NoStop}%
\bibitem [{\citenamefont {A.~K.~Tagantsev}\ and\ \citenamefont
  {Gerra}(2006)}]{Tagantsev2006}%
  \BibitemOpen
  \bibfield  {author} {\bibinfo {author} {\bibfnamefont {.}~\bibnamefont
  {A.~K.~Tagantsev}}\ and\ \bibinfo {author} {\bibfnamefont {G.}~\bibnamefont
  {Gerra}},\ }\href@noop {} {\bibfield  {journal} {\bibinfo  {journal} {J.
  Appl. Phys.}\ }\textbf {\bibinfo {volume} {100}},\ \bibinfo {pages} {051607}
  (\bibinfo {year} {2006})}\BibitemShut {NoStop}%
\bibitem [{\citenamefont {Al-Saidi}\ and\ \citenamefont
  {Rappe}()}]{Al-Saidi2010}%
  \BibitemOpen
  \bibfield  {author} {\bibinfo {author} {\bibfnamefont {W.}~\bibnamefont
  {Al-Saidi}}\ and\ \bibinfo {author} {\bibfnamefont {A.~M.}\ \bibnamefont
  {Rappe}},\ }\href {\doibase 10.1103/PhysRevB.82.155304} {\bibinfo  {journal}
  {Physical Review B}\ ,\ \bibinfo {pages} {155304}}\BibitemShut {NoStop}%
\bibitem [{\citenamefont {Lin}\ \emph {et~al.}(2016)\citenamefont {Lin},
  \citenamefont {Chernatynskiy}, \citenamefont {Nino}, \citenamefont {Jones},
  \citenamefont {Hennig},\ and\ \citenamefont {Sinnott}}]{JJones2016}%
  \BibitemOpen
\bibfield  {journal} {  }\bibfield  {author} {\bibinfo {author} {\bibfnamefont
  {F.-Y.}\ \bibnamefont {Lin}}, \bibinfo {author} {\bibfnamefont
  {A.}~\bibnamefont {Chernatynskiy}}, \bibinfo {author} {\bibfnamefont {J.~C.}\
  \bibnamefont {Nino}}, \bibinfo {author} {\bibfnamefont {J.~L.}\ \bibnamefont
  {Jones}}, \bibinfo {author} {\bibfnamefont {R.}~\bibnamefont {Hennig}}, \
  and\ \bibinfo {author} {\bibfnamefont {S.~B.}\ \bibnamefont {Sinnott}},\
  }\href@noop {} {\bibfield  {journal} {\bibinfo  {journal} {J. Appl. Phys.}\
  }\textbf {\bibinfo {volume} {120}},\ \bibinfo {pages} {045310} (\bibinfo
  {year} {2016})}\BibitemShut {NoStop}%
\bibitem [{\citenamefont {Nozaka}\ and\ \citenamefont
  {Masuda}(2007)}]{Nozaka98}%
  \BibitemOpen
  \bibfield  {author} {\bibinfo {author} {\bibfnamefont {A.}~\bibnamefont
  {Nozaka}}\ and\ \bibinfo {author} {\bibfnamefont {Y.}~\bibnamefont
  {Masuda}},\ }\href@noop {} {\bibfield  {journal} {\bibinfo  {journal}
  {Ferroelectrics}\ }\textbf {\bibinfo {volume} {357}},\ \bibinfo {pages} {276}
  (\bibinfo {year} {2007})}\BibitemShut {NoStop}%
\bibitem [{\citenamefont {Chen}\ \emph {et~al.}(2011)\citenamefont {Chen},
  \citenamefont {Schafranek}, \citenamefont {Wu},\ and\ \citenamefont
  {Klein}}]{Chen2011}%
  \BibitemOpen
  \bibfield  {author} {\bibinfo {author} {\bibfnamefont {F.}~\bibnamefont
  {Chen}}, \bibinfo {author} {\bibfnamefont {R.}~\bibnamefont {Schafranek}},
  \bibinfo {author} {\bibfnamefont {W.}~\bibnamefont {Wu}}, \ and\ \bibinfo
  {author} {\bibfnamefont {A.}~\bibnamefont {Klein}},\ }\href@noop {}
  {\bibfield  {journal} {\bibinfo  {journal} {J. Phys. D: Appl. Phys.}\
  }\textbf {\bibinfo {volume} {44}},\ \bibinfo {pages} {155301} (\bibinfo
  {year} {2011})}\BibitemShut {NoStop}%
\bibitem [{\citenamefont {Bucur}\ \emph {et~al.}(2017)\citenamefont {Bucur},
  \citenamefont {T\^{a}ase}, \citenamefont {Abramiuc}, \citenamefont
  {G.A.Lungu}, \citenamefont {Chiril\^{a}}, \citenamefont {Apostol},
  \citenamefont {R.M.Costescu}, \citenamefont {Negrea}, \citenamefont
  {Pintilie},\ and\ \citenamefont {C.M.Teodorescu}}]{Bucur2017}%
  \BibitemOpen
  \bibfield  {author} {\bibinfo {author} {\bibfnamefont {I.}~\bibnamefont
  {Bucur}}, \bibinfo {author} {\bibfnamefont {L.}~\bibnamefont {T\^{a}ase}},
  \bibinfo {author} {\bibfnamefont {L.}~\bibnamefont {Abramiuc}}, \bibinfo
  {author} {\bibnamefont {G.A.Lungu}}, \bibinfo {author} {\bibfnamefont
  {C.}~\bibnamefont {Chiril\^{a}}}, \bibinfo {author} {\bibfnamefont
  {N.}~\bibnamefont {Apostol}}, \bibinfo {author} {\bibnamefont
  {R.M.Costescu}}, \bibinfo {author} {\bibfnamefont {R.}~\bibnamefont
  {Negrea}}, \bibinfo {author} {\bibfnamefont {L.}~\bibnamefont {Pintilie}}, \
  and\ \bibinfo {author} {\bibnamefont {C.M.Teodorescu}},\ }\href@noop {}
  {\bibfield  {journal} {\bibinfo  {journal} {Applied Surface Science}\
  }\textbf {\bibinfo {volume} {432}},\ \bibinfo {pages} {27} (\bibinfo {year}
  {2017})}\BibitemShut {NoStop}%
\bibitem [{\citenamefont {Auciello}\ \emph {et~al.}(1998)\citenamefont
  {Auciello}, \citenamefont {Scott},\ and\ \citenamefont {Ramesh}}]{Au1998}%
  \BibitemOpen
  \bibfield  {author} {\bibinfo {author} {\bibfnamefont {O.}~\bibnamefont
  {Auciello}}, \bibinfo {author} {\bibfnamefont {J.~F.}\ \bibnamefont {Scott}},
  \ and\ \bibinfo {author} {\bibfnamefont {R.}~\bibnamefont {Ramesh}},\
  }\href@noop {} {\bibfield  {journal} {\bibinfo  {journal} {Phys. Today}\
  }\textbf {\bibinfo {volume} {51}},\ \bibinfo {pages} {22} (\bibinfo {year}
  {1998})}\BibitemShut {NoStop}%
\bibitem [{\citenamefont {Noheda}\ \emph {et~al.}(1999)\citenamefont {Noheda},
  \citenamefont {Cox}, \citenamefont {Shirane}, \citenamefont {Gonzalo},
  \citenamefont {Cross},\ and\ \citenamefont {Park}}]{Noheda1999}%
  \BibitemOpen
  \bibfield  {author} {\bibinfo {author} {\bibfnamefont {B.}~\bibnamefont
  {Noheda}}, \bibinfo {author} {\bibfnamefont {D.~E.}\ \bibnamefont {Cox}},
  \bibinfo {author} {\bibfnamefont {G.}~\bibnamefont {Shirane}}, \bibinfo
  {author} {\bibfnamefont {J.~A.}\ \bibnamefont {Gonzalo}}, \bibinfo {author}
  {\bibfnamefont {L.~E.}\ \bibnamefont {Cross}}, \ and\ \bibinfo {author}
  {\bibfnamefont {S.-E.}\ \bibnamefont {Park}},\ }\href {\doibase
  10.1063/1.123756} {\bibfield  {journal} {\bibinfo  {journal} {Applied Physics
  Letters}\ }\textbf {\bibinfo {volume} {2059}},\ \bibinfo {pages} {6}
  (\bibinfo {year} {1999})}\BibitemShut {NoStop}%
\bibitem [{\citenamefont {Zhang}\ \emph {et~al.}(2014)\citenamefont {Zhang},
  \citenamefont {Yokota}, \citenamefont {Glazer}, \citenamefont {Ren},
  \citenamefont {Keen}, \citenamefont {A.~Keeble}, \citenamefont {Thomas},\
  and\ \citenamefont {Ye}}]{Glazer2014}%
  \BibitemOpen
  \bibfield  {author} {\bibinfo {author} {\bibfnamefont {N.}~\bibnamefont
  {Zhang}}, \bibinfo {author} {\bibfnamefont {H.}~\bibnamefont {Yokota}},
  \bibinfo {author} {\bibfnamefont {A.~M.}\ \bibnamefont {Glazer}}, \bibinfo
  {author} {\bibfnamefont {Z.}~\bibnamefont {Ren}}, \bibinfo {author}
  {\bibfnamefont {D.}~\bibnamefont {Keen}}, \bibinfo {author} {\bibfnamefont
  {D.~S.}\ \bibnamefont {A.~Keeble}}, \bibinfo {author} {\bibfnamefont {P.~A.}\
  \bibnamefont {Thomas}}, \ and\ \bibinfo {author} {\bibfnamefont {Z.-G.}\
  \bibnamefont {Ye}},\ }\href {\doibase 10.1038/ncomms6231} {\bibfield
  {journal} {\bibinfo  {journal} {Nat. Commun.}\ }\textbf {\bibinfo {volume}
  {5}},\ \bibinfo {pages} {6231} (\bibinfo {year} {2014})}\BibitemShut
  {NoStop}%
\bibitem [{\citenamefont {Setter}\ \emph {et~al.}(2006)\citenamefont {Setter},
  \citenamefont {Damjanovic}, \citenamefont {Eng}, \citenamefont {Fox},
  \citenamefont {Gevorgian}, \citenamefont {Hong}, \citenamefont {Kingon},
  \citenamefont {Kohlstedt}, \citenamefont {Park}, \citenamefont {Stephenson},
  \citenamefont {Stolitchnov}, \citenamefont {Taganstev}, \citenamefont
  {Taylor}, \citenamefont {Yamada},\ and\ \citenamefont
  {Streiffer}}]{Setter2006}%
  \BibitemOpen
  \bibfield  {author} {\bibinfo {author} {\bibfnamefont {N.}~\bibnamefont
  {Setter}}, \bibinfo {author} {\bibfnamefont {D.}~\bibnamefont {Damjanovic}},
  \bibinfo {author} {\bibfnamefont {L.}~\bibnamefont {Eng}}, \bibinfo {author}
  {\bibfnamefont {G.}~\bibnamefont {Fox}}, \bibinfo {author} {\bibfnamefont
  {S.}~\bibnamefont {Gevorgian}}, \bibinfo {author} {\bibfnamefont
  {S.}~\bibnamefont {Hong}}, \bibinfo {author} {\bibfnamefont {A.}~\bibnamefont
  {Kingon}}, \bibinfo {author} {\bibfnamefont {H.}~\bibnamefont {Kohlstedt}},
  \bibinfo {author} {\bibfnamefont {N.~Y.}\ \bibnamefont {Park}}, \bibinfo
  {author} {\bibfnamefont {G.~B.}\ \bibnamefont {Stephenson}}, \bibinfo
  {author} {\bibfnamefont {I.}~\bibnamefont {Stolitchnov}}, \bibinfo {author}
  {\bibfnamefont {A.~K.}\ \bibnamefont {Taganstev}}, \bibinfo {author}
  {\bibfnamefont {D.~V.}\ \bibnamefont {Taylor}}, \bibinfo {author}
  {\bibfnamefont {T.}~\bibnamefont {Yamada}}, \ and\ \bibinfo {author}
  {\bibfnamefont {S.}~\bibnamefont {Streiffer}},\ }\href@noop {} {\bibfield
  {journal} {\bibinfo  {journal} {J. Appl. Phys.}\ }\textbf {\bibinfo {volume}
  {100(5)}},\ \bibinfo {pages} {051606} (\bibinfo {year} {2006})}\BibitemShut
  {NoStop}%
\bibitem [{\citenamefont {Lou}(2009)}]{Lou2009}%
  \BibitemOpen
  \bibfield  {author} {\bibinfo {author} {\bibfnamefont {X.~J.}\ \bibnamefont
  {Lou}},\ }\href@noop {} {\bibfield  {journal} {\bibinfo  {journal} {J. Appl.
  Phys.}\ }\textbf {\bibinfo {volume} {105}},\ \bibinfo {pages} {024101}
  (\bibinfo {year} {2009})}\BibitemShut {NoStop}%
\bibitem [{\citenamefont {Tagantsev}\ \emph {et~al.}(2001)\citenamefont
  {Tagantsev}, \citenamefont {Stolichnov}, \citenamefont {Colla},\ and\
  \citenamefont {Setter}}]{Tagantsev2001}%
  \BibitemOpen
  \bibfield  {author} {\bibinfo {author} {\bibfnamefont {A.~K.}\ \bibnamefont
  {Tagantsev}}, \bibinfo {author} {\bibfnamefont {I.}~\bibnamefont
  {Stolichnov}}, \bibinfo {author} {\bibfnamefont {E.~L.}\ \bibnamefont
  {Colla}}, \ and\ \bibinfo {author} {\bibfnamefont {N.}~\bibnamefont
  {Setter}},\ }\href@noop {} {\bibfield  {journal} {\bibinfo  {journal} {J.
  Appl. Phys.}\ }\textbf {\bibinfo {volume} {90}},\ \bibinfo {pages} {1387}
  (\bibinfo {year} {2001})}\BibitemShut {NoStop}%
\bibitem [{\citenamefont {Han}\ \emph {et~al.}(2010)\citenamefont {Han},
  \citenamefont {Park}, \citenamefont {Baik}, \citenamefont {Lee},
  \citenamefont {Alexe}, \citenamefont {Hesse},\ and\ \citenamefont
  {Gosele}}]{Han2010}%
  \BibitemOpen
  \bibfield  {author} {\bibinfo {author} {\bibfnamefont {H.}~\bibnamefont
  {Han}}, \bibinfo {author} {\bibfnamefont {Y.~J.}\ \bibnamefont {Park}},
  \bibinfo {author} {\bibfnamefont {S.}~\bibnamefont {Baik}}, \bibinfo {author}
  {\bibfnamefont {W.}~\bibnamefont {Lee}}, \bibinfo {author} {\bibfnamefont
  {M.}~\bibnamefont {Alexe}}, \bibinfo {author} {\bibfnamefont
  {D.}~\bibnamefont {Hesse}}, \ and\ \bibinfo {author} {\bibfnamefont
  {U.}~\bibnamefont {Gosele}},\ }\href@noop {} {\bibfield  {journal} {\bibinfo
  {journal} {Journal of Applied Physics}\ }\textbf {\bibinfo {volume} {108}},\
  \bibinfo {pages} {044102} (\bibinfo {year} {2010})}\BibitemShut {NoStop}%
\bibitem [{\citenamefont {Clark}\ \emph {et~al.}(2005)\citenamefont {Clark},
  \citenamefont {Segall}, \citenamefont {Pickard}, \citenamefont {Hasnip},
  \citenamefont {Probert}, \citenamefont {Refson},\ and\ \citenamefont
  {Payne}}]{CASTEP}%
  \BibitemOpen
  \bibfield  {author} {\bibinfo {author} {\bibfnamefont {S.~J.}\ \bibnamefont
  {Clark}}, \bibinfo {author} {\bibfnamefont {M.~D.}\ \bibnamefont {Segall}},
  \bibinfo {author} {\bibfnamefont {C.~J.}\ \bibnamefont {Pickard}}, \bibinfo
  {author} {\bibfnamefont {P.~J.}\ \bibnamefont {Hasnip}}, \bibinfo {author}
  {\bibfnamefont {M.~J.}\ \bibnamefont {Probert}}, \bibinfo {author}
  {\bibfnamefont {K.}~\bibnamefont {Refson}}, \ and\ \bibinfo {author}
  {\bibfnamefont {M.~C.}\ \bibnamefont {Payne}},\ }\href@noop {} {\bibfield
  {journal} {\bibinfo  {journal} {Zeitschrift fuer Kristallographie}\ }\textbf
  {\bibinfo {volume} {220(5-6)}},\ \bibinfo {pages} {567} (\bibinfo {year}
  {2005})}\BibitemShut {NoStop}%
\bibitem [{\citenamefont {Vanderbilt}(1990)}]{Vanderbilt_PRB_1990}%
  \BibitemOpen
  \bibfield  {author} {\bibinfo {author} {\bibfnamefont {D.}~\bibnamefont
  {Vanderbilt}},\ }\href@noop {} {\bibfield  {journal} {\bibinfo  {journal}
  {Phys. Rev. B}\ }\textbf {\bibinfo {volume} {41}},\ \bibinfo {pages} {7892}
  (\bibinfo {year} {1990})}\BibitemShut {NoStop}%
\bibitem [{\citenamefont {Wu}\ and\ \citenamefont
  {Cohen}(2006)}]{Wu_Cohen_GGA}%
  \BibitemOpen
  \bibfield  {author} {\bibinfo {author} {\bibfnamefont {Z.}~\bibnamefont
  {Wu}}\ and\ \bibinfo {author} {\bibfnamefont {R.~E.}\ \bibnamefont {Cohen}},\
  }\href@noop {} {\bibfield  {journal} {\bibinfo  {journal} {Phys. Rev. B}\
  }\textbf {\bibinfo {volume} {73}},\ \bibinfo {pages} {135116} (\bibinfo
  {year} {2006})}\BibitemShut {NoStop}%
\bibitem [{\citenamefont {Bogdanov}\ \emph {et~al.}(2016)\citenamefont
  {Bogdanov}, \citenamefont {Mysovsky}, \citenamefont {Pickard},\ and\
  \citenamefont {Kimmel}}]{Bogdanov2016}%
  \BibitemOpen
  \bibfield  {author} {\bibinfo {author} {\bibfnamefont {A.}~\bibnamefont
  {Bogdanov}}, \bibinfo {author} {\bibfnamefont {A.}~\bibnamefont {Mysovsky}},
  \bibinfo {author} {\bibfnamefont {C.~J.}\ \bibnamefont {Pickard}}, \ and\
  \bibinfo {author} {\bibfnamefont {A.~V.}\ \bibnamefont {Kimmel}},\
  }\href@noop {} {\bibfield  {journal} {\bibinfo  {journal} {PCCP}\ }\textbf
  {\bibinfo {volume} {18}},\ \bibinfo {pages} {28316} (\bibinfo {year}
  {2016})}\BibitemShut {NoStop}%
\bibitem [{\citenamefont {Stengel}\ \emph {et~al.}(2011)\citenamefont
  {Stengel}, \citenamefont {Aguado-Puente}, \citenamefont {Spaldin},\ and\
  \citenamefont {Junquera}}]{Stengel2011}%
  \BibitemOpen
  \bibfield  {author} {\bibinfo {author} {\bibfnamefont {M.}~\bibnamefont
  {Stengel}}, \bibinfo {author} {\bibfnamefont {P.}~\bibnamefont
  {Aguado-Puente}}, \bibinfo {author} {\bibfnamefont {N.~A.}\ \bibnamefont
  {Spaldin}}, \ and\ \bibinfo {author} {\bibfnamefont {J.}~\bibnamefont
  {Junquera}},\ }\href {\doibase 10.1103/PhysRevB.83.235112} {\bibfield
  {journal} {\bibinfo  {journal} {Phys. Rev. B}\ }\textbf {\bibinfo {volume}
  {83}},\ \bibinfo {pages} {235112} (\bibinfo {year} {2011})}\BibitemShut
  {NoStop}%
\bibitem [{\citenamefont {Pontes}\ \emph {et~al.}()\citenamefont {Pontes},
  \citenamefont {Garcia}, \citenamefont {Pontes}, \citenamefont {Beltran},
  \citenamefont {Andresa},\ and\ \citenamefont {Longo}}]{Pontes2012}%
  \BibitemOpen
  \bibfield  {author} {\bibinfo {author} {\bibfnamefont {D.}~\bibnamefont
  {Pontes}}, \bibinfo {author} {\bibfnamefont {L.}~\bibnamefont {Garcia}},
  \bibinfo {author} {\bibfnamefont {F.}~\bibnamefont {Pontes}}, \bibinfo
  {author} {\bibfnamefont {A.}~\bibnamefont {Beltran}}, \bibinfo {author}
  {\bibfnamefont {J.}~\bibnamefont {Andresa}}, \ and\ \bibinfo {author}
  {\bibfnamefont {E.}~\bibnamefont {Longo}},\ }\href@noop {} {\ }\BibitemShut
  {NoStop}%
\bibitem [{\citenamefont {Stengel}\ \emph {et~al.}(2009)\citenamefont
  {Stengel}, \citenamefont {Vanderbilt},\ and\ \citenamefont
  {Spaldin}}]{Stengel2009}%
  \BibitemOpen
  \bibfield  {author} {\bibinfo {author} {\bibfnamefont {M.}~\bibnamefont
  {Stengel}}, \bibinfo {author} {\bibfnamefont {D.}~\bibnamefont {Vanderbilt}},
  \ and\ \bibinfo {author} {\bibfnamefont {N.}~\bibnamefont {Spaldin}},\ }\href
  {\doibase 10.1038/nmat2429} {\bibfield  {journal} {\bibinfo  {journal}
  {Nature materials}\ }\textbf {\bibinfo {volume} {8}},\ \bibinfo {pages} {392}
  (\bibinfo {year} {2009})}\BibitemShut {NoStop}%
\bibitem [{\citenamefont {Tang}\ \emph {et~al.}(2009)\citenamefont {Tang},
  \citenamefont {Sanville},\ and\ \citenamefont {Henkelman}}]{Bader09}%
  \BibitemOpen
  \bibfield  {author} {\bibinfo {author} {\bibfnamefont {W.}~\bibnamefont
  {Tang}}, \bibinfo {author} {\bibfnamefont {E.}~\bibnamefont {Sanville}}, \
  and\ \bibinfo {author} {\bibfnamefont {G.}~\bibnamefont {Henkelman}},\
  }\href@noop {} {\bibfield  {journal} {\bibinfo  {journal} {J. Phys.: Compute
  Mater.}\ }\textbf {\bibinfo {volume} {21}},\ \bibinfo {pages} {084204}
  (\bibinfo {year} {2009})}\BibitemShut {NoStop}%
\bibitem [{\citenamefont {Sai}\ \emph {et~al.}(2005)\citenamefont {Sai},
  \citenamefont {Kolpak},\ and\ \citenamefont {Rappe}}]{Rappe2005}%
  \BibitemOpen
  \bibfield  {author} {\bibinfo {author} {\bibfnamefont {N.}~\bibnamefont
  {Sai}}, \bibinfo {author} {\bibfnamefont {A.~M.}\ \bibnamefont {Kolpak}}, \
  and\ \bibinfo {author} {\bibfnamefont {A.~M.}\ \bibnamefont {Rappe}},\
  }\href@noop {} {\bibfield  {journal} {\bibinfo  {journal} {Phys Rev B}\
  }\textbf {\bibinfo {volume} {72}},\ \bibinfo {pages} {020101(R)} (\bibinfo
  {year} {2005})}\BibitemShut {NoStop}%
\bibitem [{\citenamefont {Polanco}\ \emph {et~al.}(2012)\citenamefont
  {Polanco}, \citenamefont {Grinberg}, \citenamefont {Kolpak}, \citenamefont
  {Levchenko}, \citenamefont {Pynn},\ and\ \citenamefont {Rappe}}]{Rappe2012}%
  \BibitemOpen
  \bibfield  {author} {\bibinfo {author} {\bibfnamefont {M.~A.~M.}\
  \bibnamefont {Polanco}}, \bibinfo {author} {\bibfnamefont {I.}~\bibnamefont
  {Grinberg}}, \bibinfo {author} {\bibfnamefont {A.~M.}\ \bibnamefont
  {Kolpak}}, \bibinfo {author} {\bibfnamefont {A.~V.}\ \bibnamefont
  {Levchenko}}, \bibinfo {author} {\bibfnamefont {C.}~\bibnamefont {Pynn}}, \
  and\ \bibinfo {author} {\bibfnamefont {A.~M.}\ \bibnamefont {Rappe}},\
  }\href@noop {} {\bibfield  {journal} {\bibinfo  {journal} {Phys. Rev. B}\
  }\textbf {\bibinfo {volume} {85}},\ \bibinfo {pages} {214107} (\bibinfo
  {year} {2012})}\BibitemShut {NoStop}%
\end{thebibliography}%

\end{document}